\documentclass[runningheads]{llncs}
\usepackage[T1]{fontenc}
\usepackage{graphicx}
\usepackage{color}
\usepackage{multirow}
\usepackage{xspace}
\usepackage{bbding} % letter
\usepackage{cite}               % order multiple entries in \cite{...}
\usepackage{url}                % allow \url in bibtex for clickable links
\usepackage{xcolor}             % color definitions, to be use for...
\usepackage[]{hyperref}         % ...clickable refs within pdf...
\usepackage{breakurl}           % break too-long urls in refs

\newcommand{\para}[1]{\vspace{2pt}\noindent\textbf{{#1. }}}
\providecommand\ignore[1]{{}}
\newcommand{\sysname}{Cabin\xspace}

\newenvironment{packeditemize}{
\begin{list}{$\bullet$}{
\setlength{\labelwidth}{8pt}
\setlength{\itemsep}{0pt}
\setlength{\leftmargin}{\labelwidth}
\addtolength{\leftmargin}{\labelsep}
\setlength{\parindent}{0pt}
\setlength{\listparindent}{\parindent}
\setlength{\parsep}{0pt}
\setlength{\topsep}{3pt}}}{\end{list}}
\hypersetup{
    colorlinks = true,
    allcolors = blue,
}

\begin{document}
\title{Cabin: Confining Untrusted Programs within Confidential VMs}

%
\titlerunning{Confining Untrusted Programs within Confidential VMs}
\author{
Benshan Mei\inst{1,2}
\and Saisai Xia\inst{1,2}
\and Wenhao Wang\inst{1,2(}\Envelope\inst{)}
\and Dongdai Lin\inst{1,2}
}
\authorrunning{Benshan Mei, Saisai Xia, et al.}
\institute{
Key Laboratory of Cyberspace Security Defense, Institute of Information Engineering, Chinese Academy of Sciences, Beijing, China
\and
School of Cyber Security, University of Chinese Academy of Sciences, Beijing, China
}
\maketitle              % typeset the header of the contribution

\begin{abstract}

% This document is a model and instructions for \LaTeX.
% This and the IEEEtran.cls file define the components of your paper [title, text, heads, etc.]. *CRITICAL: Do Not Use Symbols, Special Characters, Footnotes, 
% or Math in Paper Title or Abstract.

Confidential computing safeguards sensitive computations from untrusted clouds, with Confidential Virtual Machines (CVMs) providing a secure environment for guest OS. However, CVMs often come with large and vulnerable operating system kernels, making them susceptible to attacks exploiting kernel weaknesses. The imprecise control over the read/write access in the page table has allowed attackers to exploit vulnerabilities. The lack of security hierarchy leads to insufficient separation between untrusted applications and guest OS, making the kernel susceptible to direct threats from untrusted programs. This study proposes \sysname, an isolated execution framework within guest VM utilizing the latest AMD SEV-SNP technology. \sysname shields untrusted processes to the user space of a lower  virtual machine privilege level (VMPL) by introducing a proxy-kernel between the confined processes and the guest OS. Furthermore, we propose execution protection mechanisms based on fine-gained control of VMPL privilege for vulnerable programs and the proxy-kernel to minimize the attack surface. We introduce asynchronous forwarding mechanism and anonymous memory management to reduce the performance impact. The evaluation results show that the Cabin framework incurs a modest overhead (5\% on average) on Nbench and WolfSSL benchmarks.

\keywords{
Confidential Computing
\and Trusted Execution Environment
\and Encrypted Virtualization 
\and Execution-Only Memory
\and Intra-process isolation
\and Syscall Filtering
}
\end{abstract}
\section{Introduction}
Privilege separation involves dividing privileges among different entities or processes within a system to limit potential damage caused by a compromised component. In traditional computing systems, privilege separation is achieved by separating the kernel code and userspace code. The kernel, trusted with access to all resources, is segregated from userspace programs, which are confined to their own address spaces. This separation is enforced by the CPU's execution mode and security checks performed by the memory management unit (MMU).

However, traditional privilege separation has certain drawbacks. Firstly, the kernel-user interface, represented by system calls, can allow untrusted processes to bypass kernel protections due to the large code base of the kernel. While measures like sandboxing and system call filtering can restrict attackers' ability to abuse the interface, they also increase the kernel's attack surface since these countermeasures are often implemented as part of the kernel itself. Secondly, the MMU lacks fine-grained protection for applications. The access permissions defined in the page table entries (PTEs) can only be configured as either writable or non-writable, invariably remaining readable. 
This limitation hinders the efficient implementation of execute-only memory (XOM), which is known to be effective in thwarting code-reuse attacks by making it challenging for attackers to identify usable gadgets.

In recent years, hardware-based trusted execution environment technologies, such as Intel SGX~\cite{sgxdeveloperguide}, AMD SEV~\cite{sev2020strengthening}, Intel TDX~\cite{cheng2023intel}, and ARM CCA~\cite{armcca}, have paved the way for the emergence of confidential computing. This new computing paradigm focuses on safeguarding the guest or enclave from attacks originating from potentially untrusted hosts. In the context of confidential computing, protecting the guest kernel assumes even greater significance, as it is responsible for securing users' most sensitive data. If the guest kernel is compromised, the entire CVM is at risk of compromise, potentially resulting in the leakage of any associated sensitive data. 

To address the concerns mentioned above, particularly the risks associated with CVMs, we have introduced Cabin, a novel secure execution framework tailored to confine vulnerable processes running within a CVM. Our framework leverages hardware-based isolation mechanisms, i.e., VMPL within AMD SEV-SNP, to establish a secure environment for executing vulnerable processes. Notably, with VMPL, one can assign read, write and execute permissions independently, allowing XOM to work efficiently.
Specifically, in our framework, untrusted programs are placed at a lower VMPL, ensuring the protection of the guest OS from vulnerable or malicious applications. A trusted proxy kernel within the lower VMPL acts as an intermediary, facilitating communication between confined processes and the trusted guest OS. To minimize the overhead of VMPL switches, we have designed an asynchronous method for handling events triggered by the application, such as system calls, page faults, interrupts, and exceptions. This approach reduces the number of required VMPL switches and improves overall efficiency.
Additionally, our framework allows for flexible monitoring and tracing of processes running within the user-space of the lower VMPL without requiring intervention from the guest OS. This enables the CVM owner to define custom policies for monitoring confined processes.
Lastly, our framework incorporates monitoring and logging capabilities to detect any suspicious activities and provide valuable insights into potential threats. This additional layer of security enables proactive threat detection and response. 

We have implemented a prototype of the Cabin framework on commodity AMD SEV-SNP servers, utilizing the system to provide execution protection and syscall filtering. Through evaluations on various benchmarks, including syscall routing, page fault handling, Nbench, and WolfSSL,
%, and Lighttpd, 
we observed that despite that the VMPL switch is costly (in particular, syscall routing is about several times slower than the baseline), Cabin introduces acceptable overhead in real world applications -- approximately 5\% and 10\% for the Nbench and WolfSSL benchmarks respectively.
%on computation and memory-intensive workloads. However, for I/O-intensive workloads, there is a slightly higher overhead of about xx\% due to the VMPL switch costs.
Overall, our confined secure execution framework provides a pratical solution for enhancing the security of CVMs, ensuring the protection of sensitive data from unauthorized access.

% 提出隔离执行框架将用户进程下放较低权限的用户态，增强与guest OS之间的隔离；
%To address above concerns, we proposed a novel confined secure execution framework for vulnerable process running in CVM. Our framework utilizes hardware-based isolation mechanisms to create a secure environment for executing vulnerable processes within a guest VM. By confining the execution of these processes and restricting their access to sensitive resources, we can prevent potential security breaches and protect the confidentiality of the system. Additionally, our framework incorporates monitoring and logging capabilities to detect any suspicious activities and provide insights into potential threats. Overall, our confined secure execution framework offers a robust solution for enhancing the security of virtualized environments and safeguarding sensitive data from unauthorized access.

\para{Contributions}
The contributions of this paper are as follows.
\begin{packeditemize}
    \item Designing and implementing a secure execution framework for processes within CVMs based on the fine-grained control of VMPL privilege, protecting the guest OS from direct threats posed by vulnerable or malicious programs.

    \item We propose VMPL-enhanced cross-layer execute-only protection for vulnerable programs and proxy-kernel running in lower VMPL, making it harder to find exploitable gadgets.

    \item We introduce asynchronous forwarding mechanism to minimize the performance impact on confined processes. Self-managed memory provided by the proxy-kernel further reduces the performance impact.

    \item We evaluate the performance impact of the \sysname framework on the Nbench and WolfSSL benchmarks. The evaluation results demonstrate modest overhead of the proposed framework. 
\end{packeditemize}

% \para{Road map}
% This paper is structured as follows. Following an introduction to the AMD SEV-SNP, execute-only memory, syscall filtering, and the threat model in Section \ref{sec:background}, we delve into the \sysname system design in Section \ref{sec:design}. The case studies is introduced in Section \ref{sec:case-study}. The implementation details are elaborated in Section \ref{sec:implementation}. Next, we evaluate the performance of the \sysname in Section \ref{sec:evaluation}. The related work is summarized in Section \ref{sec:related-work}. Lastly, the conclusion is drawn in Section \ref{sec:conclusion}.
\section{Background}\label{sec:background}
The emergence of new hardware-based privilege separation mechanisms within CVM presents new opportunities to enhance system and application security. With advancements in research on execute-only protection, intra-process isolation, and syscall filtering, we strive to leverage these technologies to further strengthen the system security. Therefore, we adhere to the traditional paradigm of software security, which emphasizes protecting the guest OS from potential threats posed by untrusted programs.

\subsection{SEV-SNP and VMPL}
It is crucial to protect the guest VM from malicious host in confidential computing. AMD SEV (Secure Encrypted Virtualization) is the first generation of hardware-assisted virtualization technology that solves the problem with memory encryption and isolation enhanced security~\cite{mattioli2021rome}. To defend against malicious hypervisors, the SEV and SEV-ES (Encrypted State) are proposed in succession by AMD to encrypt the memory pages and the private register contents of VMs with different keys~\cite{mofrad2018comparison}. However, the nested paging is still in the control of the hypervisor, so the SEV VM's pages could be mapped to another VM or the hypervisor~\cite{qin2023protecting}. Although the private status and pages of VM is encrypted under different keys, SEV/SEV-ES lacks integrity protection, e.g., the hypervisor can perform memory replay attacks.

In 2020, AMD introduced SEV-SNP (Secure Nested Paging), further enhancing the protection for CVM from malicious hypervisor~\cite{sev2020strengthening}. In SEV-SNP, an encrypted physical page can not be mapped to multiple owners by a malicious hypervisor. This mechanism is realized by the introduction of a Reverse Mapping Table (RMP). The RMP is a metadata table managed by the AMD Platform Security Processor (AMD PSP). It records the ownership of each system physical page and dictates read, write and execute permissions for each VMPL. On every nested-page table walking, the RMP is consulted for the permission and ownership of each system physical memory page. A nested page fault (\#NPF) will be raised on illegal access to physical pages. It is captured and handled by the hypervisor. The hypervisor manages VM Saved Areas (VMSAs) corresponding to four VMPLs. The access permission to the physical memory pages is restricted by configuring the VMPL of each page in the RMP. A vCPU can run in different VMPL contexts by switching the corresponding VMSAs with the help of the hypervisor.

Compared to page table protection, the RMP managed VMPL privilege is more flexible. Traditionally, we have NX, R/W, and U/S bits to denote non-executable, read-only, and user pages. However, the read and write permissions are not orthogonal in the page table. The RMP therefore separates the read and write access to guest physical pages, allowing one-way information flow between different VMPLs. Moreover, it separates the user and supervisor execution privilege for guest physical pages, preventing the code regions from being executed by unauthorized supervisor or user applications running in the lower VMPL. It is complementary to traditional SMEP (Supervisor Memory Execution Prevention) mechanism on x86 platform, combined with U/S and NX bits. The fine-grained privilege separation allows for strong execution protection.

\subsection{Execute-only memory}
Over the last thirty years, there has been substantial advancement in software attack and defense technology. The memory safety issue has been a long standing unsolved problem. Strategies like address space layout randomization (ASLR), stack canaries, and data execution prevention (DEP) have been used to address memory safety weaknesses. Despite these improvements, attackers persist in discovering new methods to exploit software vulnerabilities, underscoring the ongoing competition between attackers and defenders in the cyber-security realm.

The absence of code confidentiality enables attacker to gain arbitrary access to a running process by analyzing the code region for exploitable gadgets resides in the vulnerable software~\cite{zhang2017protecting}. Various software and hardware mitigation have been proposed to enhance the code confidentiality through eXecute-Only Memory (XOM)~\cite{kwon2019uxom}. XOM stands out as a straightforward and effective method that minimizes the attack surface and significantly raises the bar for attackers seeking to exploit software vulnerabilities. Through restricting access to code regions during runtime, XOM offers an additional security layer that prevents unauthorized access and manipulation of critical processes.

Previous researches have demonstrated the effectiveness XOM in strengthening software security~\cite{wang2020secure}. By preventing access to code pages, attackers are hard to find gadgets for subsequent attacks. Numerous Protection Key Registers User-space (PKRU)-based sandbox frameworks have emerged recently~\cite{vahldiek2019erim,schrammel2020donky,hedayati2019hodor}. However, due to the unprivileged nature of these hardware-based intra-process isolation mechanisms, they can be easily circumvented by exploiting the confused-deputy of the virtual-memory related syscalls~\cite{schrammel2022jenny}. Despite efforts to bolster the isolation between trusted and untrusted components, it is still considered to be weak in security-sensitive environments. Essentially, this mechanism offers safety rather than security.

Traditional page table-based memory protection is inadequate due to the absence of read/write access separation. The R/W bit on the x86 platform cannot be used to enable execute-only memory for vulnerable programs, allowing attackers to easily locate gadgets and compromise the software system in either the kernel or user-space. Even with PKRU-based execute-only protection, where read and write permissions are separated for each memory domain, it remains coarse-grained and can be circumvented in user-space~\cite{park2019libmpk}.

\subsection{Syscall filtering}
Syscall filtering plays a vital role in safeguarding OS from vulnerable and malicious software~\cite{demarinis2020sysfilter,gaidis2023sysxchg,rajagopalan2023syspart}. Existing syscall filtering mechanisms often reside in the kernel space. Once bypassed, the entire system is in danger. The PKRU-based in-process sandboxes is lightweight and efficient in ensuring the security of software~\cite{kirth2022pkru,wang2020secure}. However, the non-privileged hardware intra-process isolation primitives can be easily bypassed through the confused deputy of the syscall~\cite{park2019libmpk,connor2020pku}. In recent years, syscall filtering is widely used to ensure the security of such hardware-based intra-process isolation mechanisms~\cite{schrammel2020donky,hedayati2019hodor,schrammel2022jenny}. However, most of these syscall filtering mechanism are within the kernel space. Once compromised, the entire system is in dangers. The lack of layered defence poses a great threat to the kernel. We argue that the user and supervisor separation is insufficient and exploitable. To reduce the attack surface, the untrusted programs should be isolated from direct interact to the guest OS within CVM.

\subsection{Threat model}\label{sec:threat_model}
Our threat model aligns with that of confidential computing, where everything outside the virtual machine is considered untrusted. This includes the host OS. Our system relies on critical services from the guest OS, which is trusted. The proxy-kernel acts as a bridge between the confined processes and the guest OS, and it is also trusted. We assume that the applications are untrusted and may contain memory safety errors. Additionally, side channels and hardware attacks are outside the scope of our considerations. We operate under the assumption that the hardware functions as described in the official documentation. Furthermore, memory encryption and integrity protection measures are in place to provide an extra layer of security.
\section{Design}\label{sec:design}
We observe that the precise control over VMPL privilege on each guest physical page enables the execution of programs under a lower VMPL. However, merely possessing this control is insufficient to propose a secure isolated execution framework. The introduction of four permission bits in the VMPL mechanism addresses issues associated with traditional page table protection flags and is specifically tailored to ensuring the security of code running in the user and kernel space of the lower VMPLs. Therefore, to safeguard the guest OS, we introduce the \sysname framework, which confines untrusted programs to the user space of lower VMPL through fine-gained VMPL privilege management. To accomplish this, the architectural design is detailed as follows.

\subsection{Overview}
\begin{figure}[!htbp]
    \centering
    \includegraphics[width=0.65\linewidth]{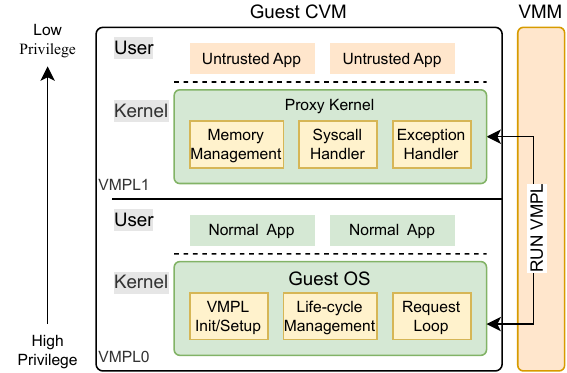}
    \caption{An overview of the \sysname framework.}
    \label{fig:cabin}
\end{figure}

Fig. \ref{fig:cabin} presents an overview of the \sysname framework. We introduce a proxy-kernel within the lower VMPL to facilitate the scheduling of processes at lower VMPLs. The proxy-kernel directly monitors confined processes and mediates the communication between the guest OS and these processes. This mediation enables the application of flexible security policies before forwarding syscalls and exceptions to the guest OS. Consequently, it establishes a layer of defense against untrusted processes. The owner of the CVM is allowed to customize policies to monitor these processes without requiring intervention from the guest OS. This design ensures the flexibility in process monitoring and tracing.
% . This mechanism allows flexible security policies to be applied before forwarding the syscalls and exceptions to the guest OS. It serves as the initial defense against untrusted processes, mitigating the direct threat from untrusted programs. 

\subsection{System design}
The \sysname shields untrusted programs in the user-space of lower VMPL. We should ensure a secure and reliable environment for untrusted applications running at lower VMPLs. Managing runtime state of confined processes is crucial. To address this, we introduce a proxy kernel to serve these confined processes. The proxy kernel functions as an intermediary between the restricted processes and the underlying guest OS, managing syscalls, and interrupts on their behalf.

The system design of the \sysname framework consists of four main components: the life-cycle management of confined threads, the context switch, syscall routing, and exception model. Below, we elaborate on each aspect of the design.

\para{Life-cycle management}
The \sysname framework supports scheduling each thread independently to the user-space of lower VMPL. The life-cycle of each thread comprises three stages: creation, entry, and exit. The guest OS manages the life-cycle of the untrusted processes as illustrated in Fig. \ref{fig:cabin}. During the initialization, the guest OS prepares the runtime environment for all lower VMPLs. Before entering the lower VMPL, the guest OS assigns a specific VMPL to each thread and synchronize the hardware state of the thread to the corresponding VMSA. Then, by requesting the hypervisor to execute in the specified VMPL, the current CPU directly switches to the corresponding VMPL and resumes the execution. Initially, Cabin enters the kernel mode of the lower VMPL, performing a series of initialization tasks for syscall and interrupt handling. Then it directly switches to the user-space, and continues the execution of the user thread. The proxy-kernel waits for syscall and interrupt events from the user-space, and forwards these events to the guest OS or handles event by itself. Upon receiving a request from the lower VMPL, the guest OS decides whether it is an interrupt or syscall event, and calls the corresponding handler in the guest OS. The request loop continues until receiving the exit and exit\_group syscalls from the lower VMPL, the guest OS no longer schedules the thread to the lower VMPL. Finally, the guest OS releases the resource for confined processes.

\para{Context switch}
The guest OS manages the context switch of confined process as usual. Compared to normal context switch in the guest OS, the hardware state of the confined process is saved in the VMSA of the lower VMPL, which is allocated by the guest OS during initialization. Because the guest OS has direct access to the hardware state of all lower VMPLs, \sysname synchronizes these state from VMSA with guest OS managed Task Control Block (TCB) on context switch. Therefore, the guest OS just loads and restores the hardware task state for confined processes at a different place. To optimize resource utilization, \sysname supports all lower VMPLs to minimize contention for limited VMSA. The confined processes are assigned to different lower VMPLs, eliminating the need to restore context when the lower VMPL is not preempted by other processes.

\para{Syscall routing}
The syscall routing logic is outlined in Fig. \ref{fig:cabin}. For confined processes running in the user space of lower VMPLs, syscalls are handled by the proxy-kernel before forwarded to the guest OS. By switching VMPL, the syscall arguments are automatically saved in the VMSA of the lower VMPL. The guest OS can directly access this hardware state. The result is returned to the proxy-kernel by modifying the VMSA of the lower VMPL. Meanwhile, certain syscalls can be directly handled by the proxy-kernel. To this end, we just simulate the syscall and sysret semantics with VMPL switching, allowing syscalls to be handled by the guest OS as usual.

\para{Exception model}
The exception in the lower VMPL should be forwarded to the guest OS in principle.  All necessary information is stored in the trap frame during a trap event, which will then be forwarded to the guest OS. Exceptions are managed in a standard manner. After handling the trap event, the guest OS requests the hypervisor to schedule the confined process. To reduce context switch, the proxy-kernel handles certain exceptions by itself. Exceptions are redirected to the guest OS as regular syscalls but are managed in a different setting. Handling exceptions involves changing the preempt mode and interrupt status of VMPL0 to ensure that the handler is invoked in a correct environment.

With the above design, we enable untrusted processes to be scheduled to the user-space of a lower VMPL, isolated from the guest OS with the VMPL hardware mechanism. The proxy-kernel mediates the communication between the untrusted programs and the guest OS. Unlike existing works, the guest OS in the \sysname framework manages all resources needed by the lower VMPLs. This innovative design brings numerous opportunities in the security aspect, which are detailed in the following sections.

\subsection{Performance optimization}
\para{Asynchronous forwarding} Most kernel operations execute quickly, rendering it costly to forward syscalls and exceptions synchronously via VMPL switching. To improve the performance, \sysname incorporates an asynchronous forwarding mechanism into the proxy-kernel. With no barriers between threads in different VMPLs, this mechanism relies on shared-memory and spinlock-based cross-thread communication. During the initialization stage, \sysname initiates a service thread that waits for requests using a spinlock. Upon entering the lower VMPL, the proxy-kernel of the lower VMPL can utilize this interface to forward syscalls and interrupts to the service thread. Once the request is completed, the proxy-kernel returns the result to the confined process, which then resumes execution until the next syscall or interrupt occurs. Compared to other asynchronous forwarding mechanisms~\cite{orenbach2017eleos,weisse2017regaining}, \sysname directly intercepts the syscall and exception in the proxy-kernel, requiring no modification to the confined programs. The untrusted programs are not allowed to directly utilize this mechanism to bypass the proxy-kernel, reducing the attack surface at the user-space.

\para{Self-managed memory}
To mitigate the performance impact of expensive VMPL switching, \sysname further incorporates anonymous memory management into the proxy-kernel, allowing direct handling of virtual memory related syscalls on anonymous pages requirement. The physical pages are granted by the guest OS, and managed by the proxy-kernel directly. When needed, the proxy-kernel requests additional memory pages from the guest OS. These pages are allocated on demand for confined processes, with any page faults on these anonymous pages being handled by the proxy-kernel, bypassing the guest OS.

% Due to implementation constraints, the proxy-kernel can only handle 4MB physical pages continuously without guest OS intervention. Currently, we limit the autonomously managed memory area to 512 MB, with GHCB and HotCalls forwarding serving as a fallback mechanism.

\section{Case studies}\label{sec:case-study}
% The case studies on the \sysname framework include three key points: execution protection for untrusted processes and proxy-kernel, syscall-filtering mechanism for untrusted processes, and exception interception for flexible process monitoring and tracing. These studies highlight potential applications of \sysname framework.
With the proxy-kernel, \sysname framework enables a series of optimization and security mechanisms for confined processes. The case studies on the \sysname framework cover three main points: execute-only protection for untrusted processes and proxy-kernel, syscalls filtering for untrusted processes, and exceptions intercepting for flexible process monitoring and tracing. These studies showcase potential applications of the \sysname framework.

\subsection{Execute-only protection}
According to the official document~\cite{sev2020strengthening}, there are four distinct permission bits for each guest physical page: read, write, user, and super execution permissions. This approach is orthogonal and distinct from traditional page table flags, where read and write access are not independent. It adds an extra layer of protection against guest physical pages. Here we present two security enhancement mechanisms based on fine-grained management of VMPL privilege.

Firstly, we propose VMPL-enhanced XOM. The guest OS revokes the read access to the code regions and then assigns execution privilege to user or super-level based on security needs. By restricting execute-only VMPL privilege to the code pages, we prevent attackers from exploiting vulnerabilities both in the kernel and user-space of lower VMPL. Due to the privileged nature of VMPL mechanism, it overcomes the short comings that arises in most non-privileged hardware-based intro-process isolation mechanism, i.e., PKRU-based XOM.

% In addition to execute-only protection, the strict separation between user and kernel space is also indispensable. However, in an environment where SMAP/SMEP is disabled, the privileged instruction can be utilized for efficient intra-process isolation ~\cite{wang2020seimi}. The separation between user and kernel-space is weaken in this scenario. By leveraging the VMPL privilege, we rebuild the separation between the kernel and user space, preventing the execution of untrusted user-space code in the kernel space of lower VMPL. This is achieved by employing the fine-grained separation of super and user execution privilege. It can be utilized as an enhanced SMEP mechanism in SMEP (Supervisor Mode Execution Prevention) disabled environment, such as Dune~\cite{belay2012dune} and SEIMI~\cite{wang2020seimi}.

Secondly, we introduce VMPL-enhanced cross-layer execute-only protection, serving as an enhanced SMEP mechanism. This is achieved through fine-grained separation of user and super execute privilege. As the VMPL further separates the execution privilege for user and kernel space, we can not only prevent the execution of untrusted user code in the kernel space, but also forbid the privileged code from being executed in user space even in the absence of U/S bit protection in the page table.

By utilizing the VMPL hardware mechanism, \sysname establishes a strict boundary between the kernel and user space at lower VMPLs. It enforces both intra-process isolation and cross-layer protection. It makes the attacker more difficult to exploit vulnerabilities at the lower VMPL. Overall, we utilizes VMPL to enable "one-way visibility" of a reference monitor, ensuring that code regions cannot be inspected and altered at the lower VMPL.

% \subsection{Intra-process isolation}
% With \sysname framework, we allow threads to run in an isolated environments, achieving strong thread-level isolation. Each thread has their own root page table at the user space of lower VMPL. The page tables are synchronized accoss VMPL. we allow the proxy-kernel to provide additional syscalls to provide intra-process isolation for confined process. Although those threads.

\subsection{Process monitoring}
Since the proxy-kernel mediates the communication between the guest OS and untrusted programs. It can directly handles the syscall and exception from user-space before forwarding to the guest OS. This mechanism can be leveraged to enhance the performance or track the execution of confined user programs.

\para{Syscall filtering}
\sysname introduces VMPL-enforced execute-only protection to reduce the attack surface for vulnerable programs. However, it is not sufficient for malware. The syscall filtering can be leveraged as a layer of defense in the lower VMPL without intervention from the guest OS.

\para{Process tracing}
By intercepting the breakpoint exception, \sysname enable dynamic monitoring of untrusted programs without guest OS intervention, offering a flexible tracing mechanism. This allows us to utilize the hardware breakpoint based dynamic intercepting mechanism without relying on the guest OS. Similar to the kprobes mechanism in the Linux kernel, we enable automatic process tracing running in the lower VMPL. Additionally, dynamic instrumentation can be readily supported on \sysname for closed-source binaries.

\para{Malware analysis}
For malware where no source code can be accessed, the exception intercepting mechanism allow flexible security policies to be applied to each confined process without requiring intervention from guest OS. It is especially useful in analyzing the behaviour of malware. Since the policies are outside of the guest OS, modifying the security policy is made simple.

% The case studies above showcase the potential of the proposed \sysname framework. However, some of the case studies are still in-progress. The self-managed virtual memory is not yet perfect and contains bugs. Therefore, we will not present the performance evaluation of these optimization.
\section{Implementation}\label{sec:implementation}
The current implementation of the \sysname framework supports Linux running on AMD SEV-SNP enabled CPUs. It is based on the lasted infrastructure from AMD SEV\footnote{\protect\url{https://github.com/AMDESE}}. To streamline the management of confined processes, \sysname consists of a kernel module and a proxy-kernel. The kernel module manages the life-cycle of the confined processes, while the proxy-kernel serves these processes in the lower VMPL. The kernel module comprises approximately 6600 lines of code (LoCs), the proxy-kernel has 11000 LoCs, and the musl-libc~\footnote{\protect\url{https://musl.libc.org}} contributes around 500 LoCs for GHCB protocol-based syscall forwarding mechanism.

% The proxy-kernel in \sysname currently operates as a LibOS. We recognize that a self-contained proxy-kernel could offer an ideal setup for \sysname. Nonetheless, both setups show no inherent disparities in security, nor impact the performance evaluation. However, providing a self-contained proxy-kernel can be a future work. 

\para{Application interface}
We offer two interfaces for applications that need confinement. The vmpl\_init is utilized to setup the environment at the process level. The vmpl\_enter\_user is used to prepare thread-level resources and enter the lower VMPL. The thread can be scheduled independently to the lower VMPL. Besides, we introduce a preload library for unmodified binary programs. There is no need to modify or statically instrument the source code, greatly reducing the deployment effort.

\subsection{Syscalls and interrupts handling}
In the \sysname framework, the proxy-kernel directly handles the syscalls and the interrupts from the confined process. The forwarding mechanism follows standard GHCB protocol~\cite{AMDSVSM-56421}. The MSR (Model-Specific Registers) protocol serves as a bootstrapping mechanism for GHCB protocol before GHCB registration. Once the GHCB is registered at the lower VMPL, \sysname directly shifts to the GHCB-based forwarding mechanism. To ensure the functionality and efficiency of syscalls and interrupts handling, the implementation of the \sysname framework includes the following features: the vDSO support, asynchronous forwarding, and transparent debugging.

\para{Syscall routing} 
\sysname supports anonymous memory management in the proxy-kernel. Certain virtual memory related syscalls, such as mmap, munmap, mprotect, and mremap can be handled by the proxy-kernel without VMPL switching. Unsupported syscalls are still forwarded to the guest OS. A simple filtering mechanism is also implemented in the forwarding logic, allowing intercept each syscalls independently with priority. To enforce syscall security, security policies can be enforced prior to entering the lower VMPL.

\para{vDSO support} The vDSO (virtual Dynamic Shared Object) is a conventional mechanism that allows programs to make syscalls directly without transition to kernel mode~\cite{vdso}. It is a memory area used by the kernel to provide optimized versions of commonly used syscalls (i.e., clock\_gettime). This improves performance by reducing the overhead of context switch. The vDSO is mapped into the address space of every user-space process, allowing programs to access it easily when making these syscalls. \sysname naturally supports such mechanism by allowing access to those memory pages at lower VMPL.

\para{Asynchronous forwarding}
To reduce the costly VMPL switching, the asynchronous forwarding mechanism is derived from SGX-HotCalls~\cite{weisse2017regaining}. By removing the Intel SGX-related components, it seamlessly integrates with the \sysname framework. Unlike the original version, this mechanism is integrated into the syscall and interrupt handlers of the proxy-kernel. Currently, the \sysname framework supports asynchronous forwarding for syscalls, while exceptions and interrupts remain GHCB protocol-based synchronous forwarding mechanism.

\para{Transparent debugging}
Transparent debugging is essential in the \sysname framework for confined processes. It ensures seamless debugging capabilities for the lower VMPL. The hardware state of the lower VMPL is synchronized with the guest OS-managed TCB, encompassing debug registers, during context switches. The trap frame from the user-space of lower VMPLs is delivered to the guest OS to facilitate the handling of breakpoints and debug exceptions triggered at the lower VMPL, allowing transparent debugging for confined processes.

\subsection{Dynamic VMPL management}
It is crucial to adjust the VMPL permission of each physical memory pages for a confined process to run in the user space of the lower VMPL on AMD SEV-SNP platform. This process mainly includes intercepting syscalls and exceptions, as outlined below.

\begin{table}[!htbp]
    \vspace{-12pt}
    \centering
    \caption{System call categories.}
    \label{tab:system_call_categories}
    \begin{tabular}{|c|c|}
    \hline
    % \multicolumn{2}{|c|}{\textbf{Table III: System Call Categories}} \\ \hline
    \textbf{Category} & \textbf{syscalls} \\ \hline
    \multirow{4}{*}{Virtual Memory} & mmap, mremap, munmap, brk, \\
                                    & mprotect, pkey\_mprotect, madvise, \\
                                    & shmat, shmdt, remap\_file\_paegs, \\
                                    & mlock, mlock2, mlockall \\ \hline
    \end{tabular}
    \vspace{-12pt}
\end{table}

\para{Syscall interposition}
To update VMPL permission on time, we adjust the permission of the relevant physical pages after each system call and page fault, so that the process can be running at a lower VMPL. We identified several virtual memory-related syscalls (e.g., brk, mmap) in Table. \ref{tab:system_call_categories}. However, these syscalls do not always populate the page table due to lazy allocation. To streamline the process, we still traverse the page table and grant access to corresponding memory area. Despite being imprecise and inefficient, the evaluation shows a modest overhead through other optimizations.

The syscalls mentioned above typically accept memory address and length as arguments and can be easily monitored for VMPL management. However, certain other syscalls (e.g., read) implicitly alter the page tables by synchronizing memory contents between hardware storage and memory. In these cases, the kernel will inform subscribers before and after modifying the page table. We utilize such notification mechanism to adjust the VMPL permission.

\para{Page fault interception}
Due to the lazy-allocation and demand paging mechanism, we update the corresponding VMPL permission after the guest OS successfully handles the page fault on non-present and copy-on-write (COW) pages. However, it is not sufficient to run the process at lower VMPL. The kernel pre-allocates physical pages before actually accessing those pages due to the prefault mechanism. Therefore, we promptly adjust the VMPL permission for prefault pages. Otherwise, it may cause RMP permission violations caused by being unable to access these physical pages at a lower VMPL.

Notably, a better way to improve the performance is to grant the entire memory access rights to all lower VMPLs according to the firmware specification~\cite{AMDABI-56860}. In this way, all guest physical pages are allowed to access at lower VMPL. However, it is still necessary to conditionally adjust the VMPL permissions for security. It is a complex task to track all updates to the page table of a process. Our prototype focuses on demonstrating the viability and security of running untrusted programs within the user-space of a lower VMPL. Therefore, we do not focus on a precise tracking mechanism in this work. However, a precise page table tracking mechanism can be realized with further efforts.
\section{Performance Evaluation}\label{sec:evaluation}
In this section, we evaluate the performance of the \sysname framework. The evaluation is performed in a single-threaded environment. This includes using GHCB protocol and HotCalls to forward syscalls and page faults to the guest OS. Afterwards, we measure the performance on Nbench and WolfSSL benchmarks. The evaluation is performed on a dual-socket 3rd Gen AMD EPYC processor (code-named Milan) with 128 logical cores and 64GB RAM, supporting the SEV-SNP technology. The host system operates QEMU 6.1.50 on Ubuntu 22.04 (kernel version 6.5.0-rc2-snp-host), while the VM is allocated with 64 vCPUs and 16GB RAM, running Ubuntu 22.04 (kernel version 6.5.0-snp-guest).

\begin{figure}[!htbp]
    \vspace{-12pt}
    \centering
    \includegraphics[width=\linewidth]{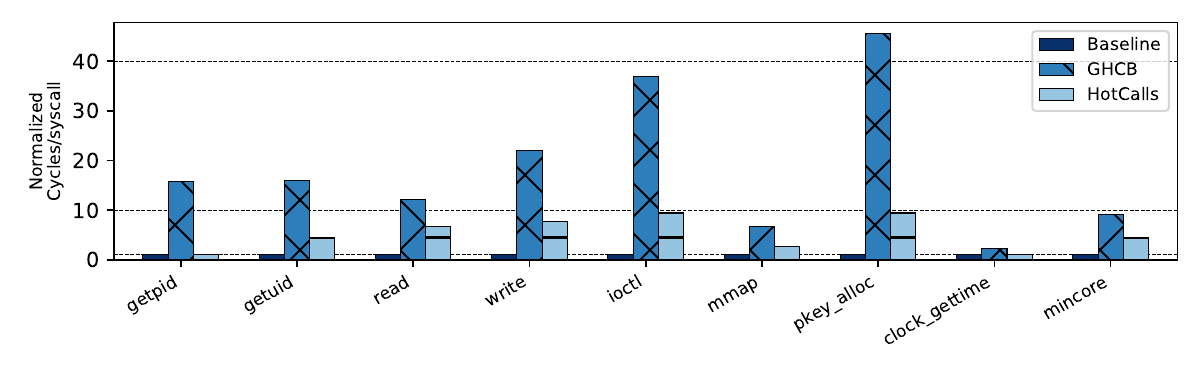}
    \caption{The evaluation of syscall overhead in different scenarios.}
    \label{fig:syscall}
    \vspace{-12pt}
\end{figure}

% Table \ref{tab:syscall} illustrates the time taken to execute each syscall 10,000 times under different scenarios. The GHCB/MSR protocol-based forwarding mechanism requires significant time consumption than original syscall instruction. The dynamic VMPL management mechanism introduces significant overhead on read, mmap, and munmap syscalls. By employing the HotCalls mechanism for syscall forwarding, a noticeable reduction in execution time is observed compared to the GHCB/MSR protocol. However, HotCalls still falls short in speed compared to the original syscall method. This is attributed to its asynchronous nature, leading to varying latency across syscalls. Notably, as \sysname supports the vDSO mechanism, there is no impact on the traditional clock\_gettime syscall.
\para{Syscall} Fig \ref{fig:syscall} depicts the time taken to execute each syscall 10,000 times under various conditions. The GHCB protocol-based forwarding mechanism incurs more time consumption than the original syscall instruction. Employing the HotCalls mechanism for syscall forwarding shows a noticeable reduction in execution time compared to the GHCB protocol. However, HotCalls still lags behind in speed compared to the original syscall method due to its asynchronous nature, resulting in varying latency across syscalls. Notably, the dynamic VMPL management mechanism introduces significant overhead on read and mmap syscalls. Importantly, with \sysname supporting the vDSO mechanism, there is no impact on the clock\_gettime syscall.

% Due to technical issues, some syscalls like open, close, socket, and unmap cannot currently be forwarded with HotCalls. Despite this limitation, it exhibits reasonable overhead compared to other syscalls, even with GHCB-based forwarding mechanism.

\ignore{
\begin{table}[!htbp]
    \centering
    \begin{tabular}{lrrr}
        \hline  
         &  Baseline & GHCB & HotCalls\\
         \hline
         none & 140 & 136 & 139 \\
        getpid & 788  & 12370 & 914 \\
        getuid & 742 & 11925 & 3307 \\
        read & 1335 & 16243 & 8944 \\
        write & 623 & 13717 & 4816 \\
        ioctl & 395 & 14596 & 3749 \\
        mmap & 3348 & 22765 & 8958 \\
        % mprotect & 4702 & 24197 & 31583 \\
        % unmap & 5254 & 25301 & \\
        pkey\_alloc & 360 & 16394 & 3440 \\
        clock\_gettime & 137 & 323 & 161 \\
        mincore & 1926 & 17494 & 8543 \\
        \hline
    \end{tabular}
    \caption{Delay in handling syscalls in different scenarios.}
    \label{tab:syscall}
\end{table}
}

\para{Page fault} Table \ref{tab:page_fault} presents the duration for handling page faults in different scenarios. When assigning 10000 private memory pages via mmap syscall, each memory page access prompts a page fault without preloading. Remarkably, the time taken to manage page faults is considerable, almost matching the overhead from GHCB protocol. In the lower VMPL, the page fault forwarding mechanism operates approximately three times as slowly as in the original user-space. Compared to the forwarding syscall, the page fault has a greater performance impact because it involves synchronizing the trap frame to guest OS. Forwarding certain page faults with HotCalls is possible, but the current implementation hasn't adopted a HotCalls-based forwarding mechanism.

\begin{table}[!htbp]
    \caption{Delay in handling page exception in different scenarios.}
    \label{tab:page_fault}
    \centering
    \begin{tabular}{|c|c|c|c|c|}
    \hline
            & Baseline & VMPL-CPL0 & VMPL-CPL3 \\
    \hline
    page fault & 13026 & 29627 & 29936 \\
    \hline
    \end{tabular}
    \vspace{-12pt}
\end{table}

In the following, we evaluate the impact on classical performance benchmarks, showcasing the advantages of the \sysname framework.

\ignore{
\para{SPEC 2006}
Figure \ref{fig:spec2006} shows the experimental results on spec 2006, including the performance loss with and without HotCalls. It can be seen that the performance loss is significantly reduced when HotCalls is used, indicating that HotCalls plays an important role in improving system performance.

\begin{figure}[!htbp]
    \centering
    \includegraphics[width=1.0\linewidth]{template//figures/CINT2006.001.txt-time.pdf}
    \caption{SPEC CPU2006}
    \label{fig:spec2006}
\end{figure}

\para{SPEC 2017}
Figure \ref{fig:spec2017} shows the experimental results on spec 2017, including the performance loss with and without HotCalls. It can be seen that the performance loss is significantly reduced when HotCalls is used, indicating that HotCalls plays an important role in improving system performance.

\begin{figure}[!htbp]
    \centering
    \includegraphics[width=1.0\linewidth]{template//figures/CINT2006.001.txt-time.pdf}
    \caption{SPEC CPU2017}
    \label{fig:spec2017}
\end{figure}

\para{MbedTLS}
We evaluated the performance impact of \sysname on the the MbedTLS benchmark, which covered different encryption algorithms. We can see the performance impact of \sysname on these encryption algorithm implementations, which is crucial for evaluating the performance of \sysname in real system applications. Through these tests, we can better assess the advantages and limitations of \sysname in practical applications, providing a stronger basis for system design and optimization.

\para{OSMark} 
\cite{cui2009osmark}
}

\begin{figure}[!htbp]
    \vspace{-12pt}
    \centering
    \includegraphics[width=0.75\linewidth]{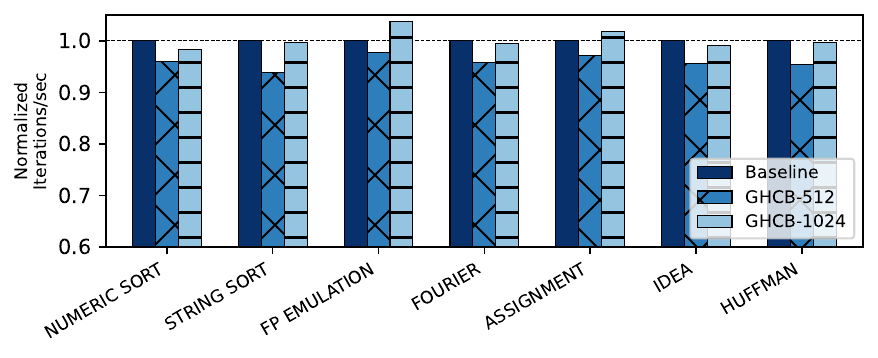}
    \caption{The performance evaluation on Nbench.}
    \label{fig:nbench}
    \vspace{-12pt}
\end{figure}

\para{Nbench~\cite{nbench}}
Fig. \ref{fig:nbench} shows the evaluation of the \sysname framework on Nbench. This benchmark includes ten calculation-intensive tasks. We utilize proxy-kernel provided mmap and munmap syscalls for small-scale anonymous memory requirement. The GHCB-512 and 1024 indicate that the proxy-kernel manages 512 and 1024 memory pages continuously without guest OS intervention. It is evident that despite implementing the self-managed memory mechanism, there is still an overall performance overhead. This is due to the necessity of forwarding all other syscalls and interrupts. Although \sysname supports vDSO based clock\_gettime, it is still forwarded to the guest OS in Nbench. Nevertheless, as the proxy-kernel manages more physical pages, the performance impact notably decreases across most benchmarks. Additionally, there is a substantial performance enhancement observed in FP EMULATION and ASSIGNMENT when the proxy-kernel manages more memory pages.

\begin{figure}[!htbp]
    \centering
    \includegraphics[width=1.0\linewidth]{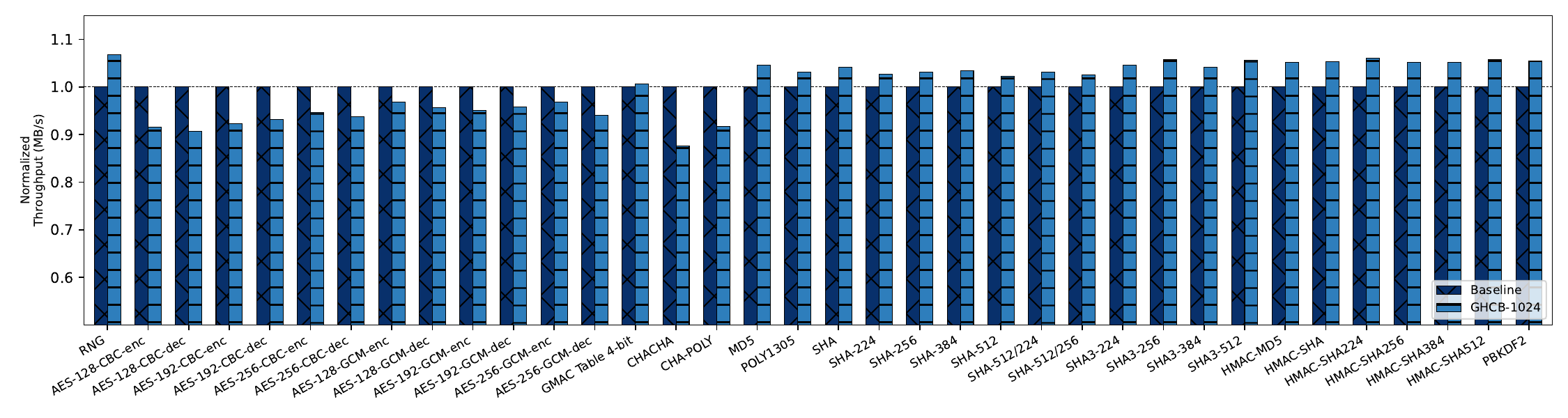}
    \caption{The performance evaluation on WolfSSL benchmark.}
    \label{fig:wolf}
\end{figure}

\para{WolfSSL~\cite{wolfssl}}
We evaluate the \sysname framework on WolfSSL benchmark. This benchmark consists of evaluation on cryptography algorithms, such as encryption, decryption, digests, and signature verification. Here, the anonymous memory allocation is also handled by the proxy-kernel rather than the guest OS. As illustrated in Figure~\ref{fig:wolf}, over a half of tasks perform significant better than baseline, while the other remains an overall performance overhead of about 1\% to 10\%. This indicates that in certain cases, using autonomous management of anonymous page memory allocation can bring performance improvements.

\vspace{-12pt}

In above evaluations, the \sysname incurs significant overhead on each syscall due to costly VMPL switching. Both syscall and exception forwarding require more cycles when the process is scheduled to the lower VMPL. However, \sysname incurs modest overhead in most cases on Nbench and WolfSSL benchmarks. The performance impact can be reduced significantly with asynchronous HotCalls mechanism and self-managed memory mechanisms, thereby outstanding the advantage of confined execution of \sysname.

\ignore{
\para{IOzone Filesystem Benchmark~\cite{iozone}}

\subsection{Macro-benchmarks}
The macro-benchmarks consists of the performance of \sysname in confined execution of varies programs.  We run the benchmarks with the following settings: in original user-space (vanilla), in the user space of lower VMPL (vmpl), in the user space of lower VMPL with hotcalls enabled (hotcalls). Compared to Nbench, these group of benchmarks are I/O intensive, requiring more syscall forwarding.

\para{Lighttpd} The lighttpd is a classical light-weight single-threaded server framework. Figure \ref{fig:lighttpd} shows the performance loss of Lighttpd on the proposed framework, including the performance loss with and without HotCalls. From the above evaluations, the \sysname incurs acceptable on most benchmarks. The performance loss of Lighttpd mainly focuses on the increase in request response time, especially under high load conditions. With HotCalls enabled, the performance loss has been significantly improved, and the request response time is noticeably reduced.

We evaluated lighttpd version 1.4.41. We evaluated its performance using http\_load~\cite{httpload}.

The measurement consisted of 100 concurrent clients connections, fetching a total of 1 million 20 KB pages. The connections were over the local loopback to maximize available link bandwidth. Unmodified lighttpd was able to serve an average of 53,400 pages per second, with average response latency of 1.52 milliseconds.

\begin{figure}[!htbp]
    \centering
    \includegraphics[width=0.75\linewidth]{template//figures/lighttpd.pdf}
    \caption{Lighttpd}
    \label{fig:lighttpd}
\end{figure}

\para{thttpd}
The thttpd~\cite{thttpd} is a classical light-weight single-threaded server framework. We evaluated its performance using http\_load~\cite{httpload}.
}
\section{Discussion}\label{sec:discussion}
Every security mechanism comes with a cost, and the security framework we propose is no exception. The advantages and limitations of the proposed \sysname framework are outlined in the following.
\subsection{Advantages}
% The advantages of the \sysname framework encompass defense in-depth, compatibility, and superior design compared to alternatives. This framework allows a layered defense to vulnerable and malicious programs and is compatible to the other enclave or sandbox frameworks.

\para{Defense in-depth}
Compared to traditional sandbox frameworks, \sysname shields untrusted processes to the lower VMPL within the same CVM, preventing vulnerabilities from malicious exploits with VMPL-enhanced execute-only memory and cross-layer execution prevention. By isolating processes at the user space of lower VMPL, \sysname provides layered protection for the guest OS within the CVM. This framework allows for flexible process monitoring and tracking of untrusted legacy applications without requiring intervention from the guest OS.

\para{Compatibility}
One advantage of the \sysname is the compatibility with other frameworks. In the Secure Virtual Machine Service Module (SVSM)~\cite{AMDSVSM-58019}, the guest OS operates in lower VMPL other than VMPL0. Our schema naturally aligns with this framework. In this case, the process is scheduled to at most two VMPLs. To accommodate other frameworks like Veil~\cite{ahmad2023veil}, \sysname require at least one VMPL lower than the guest OS. The trusted services and enclaves are positioned at higher VMPLs, while the untrusted processes are scheduled to a lower VMPL. However, Veil positions the guest OS at the lowest VMPL, rendering it challenging to integrate the \sysname framework. Cabin is also naturally supports PKRU-based sandbox frameworks~\cite{wang2020secure,kirth2022pkru,voulimeneas2022you,schrammel2022jenny}, which can still be used to enhance intra-process isolation for confined processes at lower VMPLs.

\para{Alternative design}
Compared to safeguarding the proxy-kernel with VMPL, an alternative design to the proposed execution protection mechanism is based on the SVSM framework. This approach restricts read access to the code regions of guest OS. However, there exist numerous code regions necessitating read access and even modification rights. Modifying these code regions allows the kernel to dynamically change behavior during runtime. Consequently, it is less practical than protecting a minimal proxy-kernel in lower VMPL.

\subsection{Limitations}
One drawback of the \sysname framework is the performance impact. The VMPL switching leads to delays in syscall and exception handling. The imprecise page tracking for dynamic VMPL management results in extra overhead. Currently, \sysname does not well support thread migration across CPUs. Because the GHCB is not shared among CPU cores, \sysname binds the thread to one CPU, limiting task scheduling flexibility. Other constraints involve multi-threading and multi-processing. Although \sysname supports preemptive scheduling, the incomplete support for fork and clone syscalls limits the application to single-thread environment. Nevertheless, it is possible to schedule child threads to the user space of lower VMPLs while keeping the main thread in the original user space. Most issues can be solved with further effort, but the delays from VMPL switching remain a challenge to efficiently address.

\subsection{Extending to other CVM platforms}\label{sec:extending}
Although the \sysname framework is based on the latest feature from AMD SEV-SNP, it can be extended to other CVM platforms such as Intel TDX and ARM CCA. By introducing a proxy-kernel within an isolated CVM, we shield the guest OS from potential threats posed by untrusted processes. The communication between confined processes and the guest OS is managed by the proxy-kernel and the trusted hypervisor located outside the CVM. As for the Intel TDX, the TDX Module facilitates communication between the proxy kernel and the guest OS across different CVMs. Meanwhile, in ARM CCA, the Realm Management Monitor (RMM) oversees the interaction between the proxy-kernel and the guest OS. In both scenarios, trusted hypervisors like Intel TDX Module and ARM RMM play a crucial role in establishing a secure channel between different CVMs.

Recently, ARM CCA introduced support for different planes within a CVM~\cite{armplanes}. Each plane is essentially a separate VM, with a shared guest physical address space. Plane 0 holds more privilege and can host a paravisor to control switches between planes and restrict other planes' memory access. Similarly, less privileged planes can be used to shield untrusted applications from guest OS.

\section{Related Work}\label{sec:related-work}

\para{AMD SEV-SNP and VMPL}
Various researches are underway to enhance the security of AMD SEV~\cite{li2022systematic,buhren2022resource,qin2023protecting}.
The SVSM~\cite{AMDSVSM-58019} framework leverages VMPL0 to protect secure service from untrusted guest OS. Hecate~\cite{ge2022hecate} uses VMPL0 as a trusted L1-hypervisor to facilitate communication between the guest OS and untrusted hypervisor. SVSM-vTPM~\cite{narayanan2023remote} is a security-enhanced vTPM based on the SVSM framework, leveraging VMPL0 to isolate the virtual TPM (vTPM) from the guest OS, ensuring the integrity of vTPM's functions. CoCoTPM~\cite{pecholt2022cocotpm} reduces the trust needed towards the host and hypervisor by running a vTPM in an encrypted VM using AMD SEV. Honeycomb~\cite{mai2023honeycomb} is a secure GPU computation framework that runs a validator within VMPL0, which inspects the binary code of a GPU kernel to ensure that every memory instruction in the kernel can solely reach designated virtual address space, utilizing static analysis. The mushroom~\cite{mushroom} framework runs integrity protected workloads based on AMD's SEV-SNP technology, which could be the basis of a secure remote build system. Veil~\cite{ahmad2023veil} is a service framework providing secure enclave and services for process and the guest OS respectively. In general, these works follow traditional threat model of confidential computing, and do not focus on untrusted applications in CVM, while the \sysname framework protects the guest OS by confining untrusted programs to the user space of lower VMPLs.

\para{Execute-only memory}
Execute-only memory (XOM)~\cite{chen2017norax, kwon2019uxom} is an effective method in software security. PicoXOM~\cite{shen2020fast} is an efficient XOM mechanism based on ARM's Data Watchpoint and Tracing unit for embedded systems. Nojitsu~\cite{park2020nojitsu} leverages XOM-Switch to enforce execute-only permission for static code regions in JIT. SECRET~\cite{zhang2017protecting} protects COTS binaries from disclosure-guided code reuse attacks, while MonGuard~\cite{wang2020secure} applies PKRU-based XOM protection to the multi-variant execution (MVX) monitor. IskiOS applies XOM to safeguard code pages of a unikernel~\cite{gravani2021fast}. Cerberus~\cite{voulimeneas2022you} is a notable sandbox framework that protect the reference monitor with PKRU-based XOM. To the best of our knowledge, the fine-grained control over VMPL permissions has not been utilized to enhance execute-only protection for untrusted programs in previous studies.

\para{Intro-process isolation}
The lightweight PKRU-based intra-process isolation mechanism is also a hot research topic in recent years~\cite{vahldiek2019erim,hedayati2019hodor,kirth2022pkru}.
Various research efforts have been made to enhance the security of PKRU-based isolation mechanisms~\cite{voulimeneas2022you,schrammel2022jenny}.
However, its unprivileged nature makes it susceptible to bypassing in user-space through side-effects or confused-deputy issues from syscalls~\cite{park2019libmpk}.
Attackers can exploit this vulnerability by constructing unsafe instruction sequences to gain unauthorized access to sensitive data and code~\cite{connor2020pku}. Such systems
require complex syscall filtering policy to prevent WRPKRU exploitation and enforce the security of their sandbox~\cite{schrammel2022jenny}.

\para{Syscall filtering}
Securely confining untrusted legacy applications has been a long-standing challenge for the past decades~\cite{garfinkel2004ostia,potter2007secure,ibrahim2021secure}. The syscall filtering plays a crucial role in traditional software system security~\cite{demarinis2020sysfilter,gaidis2023sysxchg,rajagopalan2023syspart}, including container security~\cite{song2023value}. The syscall filtering is also widely applied in PKRU-based intro-process isolation mechanism~\cite{schrammel2022jenny,christou2023binwrap}. PHMon~\cite{delshadtehrani2020phmon} and FlexFilt~\cite{delshadtehrani2021flexfilt} introduces new hardware design for efficient syscall filtering and process monitoring on RISC-V platform. Nevertheless, due to limited privilege separation, these mechanisms still confine to conventional user and kernel separation.
\section{Conclusion}\label{sec:conclusion}

\sysname is an isolated execution framework that effectively shields untrusted programs from guest OS within CVM. By introducing a trusted proxy-kernel for untrusted applications, \sysname enables efficient and flexible process monitoring and tracing, enhancing a layered security defense outside of the guest OS. By utilizing VMPL-enforced execute-only protection, \sysname making it harder for vulnerabilities to be exploited at lower VMPL.
With fine-grained control over VMPL execution privilege, \sysname further isolates the proxy-kernel and confined processes, strengthening the cross-layer isolation between the user and kernel space of lower VMPLs. 
To reduce the performance impact, \sysname integrates asynchronous forwarding mechanism and self-managed memory allocation in the proxy-kernel. In essence, the \sysname framework can be generalized to other commercial CVM platforms as well. The evaluation results on Nbench and WolfSSL benchmarks demonstrate modest performance overhead for confined processes.

% \sysname is an isolated execution framework that effectively shields untrusted programs from the guest OS within CVM. By introducing a trusted proxy-kernel for untrusted applications, \sysname enables efficient and flexible process monitoring and tracing, enhancing security outside of the guest OS. Through the use of VMPL-enforced execute-only protection, \sysname makes it more challenging for vulnerabilities to be exploited at lower VMPL levels. 
% With fine-grained control over VMPL execution privilege, \sysname further isolates the proxy-kernel and confined processes, strengthening the cross-layer isolation between the user and kernel space of lower VMPLs. To minimize the performance impact, \sysname integrates an asynchronous forwarding mechanism and self-managed memory allocation in the proxy-kernel. Evaluation results on Nbench and WolfSSL benchmarks show a modest performance overhead for confined processes.

%\subsubsection{Acknowledgements} Please place your acknowledgments at
%the end of the paper, preceded by an unnumbered run-in heading (i.e.
%3rd-level heading).
\para{Acknowledgment} This work was supported by National Natural Science Foundation of China (Grant No.62272452). Corresponding author: Wenhao Wang (\href{mailto:wangwenhao@iie.ac.cn}{wangwenhao@iie.ac.cn}).

\bibliographystyle{splncs04}
\bibliography{Biblio-Bibtex}
\end{document}